# Surface oxidation of Pt and PtPd alloys during NO conversion investigated by Atom Probe Tomography

Yoonhee Lee[1,3]*, Daniel Dobesch [2], Patrick Stender [1], Ute Tuttlies[2], Ulrich Nieken[2], Guido Schmitz [1]

[1] *Institute of Materials Science, University of Stuttgart, Heisenbergstr. 3, 70569 Stuttgart, Germany*

[2] *Institute of Chemical Process Engineering, University of Stuttgart, Böblinger Str. 78, 70199 Stuttgart, Germany*

[3] *Max Planck Institute for Sustainable Materials, Max-Planck-Straße 1, 40237 Düsseldorf, Germany*

*y.lee@mpi-susmat.de

## Abstract

In this study, the dynamic oxidation state changes of pure Pt and 50 at% Pt–Pd alloy catalysts were investigated during a temperature ramp from 80 to 450 °C under a reactive gas mixture of 500 ppm NO and 3% $O_2$ in $N_2$. These changes are closely correlated with variations in NO conversion efficiency. Sharp-tip specimens were prepared from Pt and Pt–Pd alloy wires via electrochemical polishing and Focused Ion Beam (FIB) annular milling, with the hemispherical tip apex serving as a model for nanoscale catalyst surfaces. The samples were exposed to the reactive atmosphere in a dedicated reaction chamber and subsequently analyzed using atom probe tomography (APT). Effective oxide thicknesses and three-dimensional surface morphology were quantitatively evaluated. A pronounced decrease of this thickness was observed during the first cooling and second heating cycles, particularly below 200 °C, indicating a reversible redox behavior of both Pt and Pt–Pd alloy surfaces. This is related to the inverse hysteresis of NO conversion measured for both systems. The redox reversibility is attributed primarily to reaction kinetics rather than thermodynamic stability.

1. Introduction

In the lean gas environment of a Diesel Oxidation Catalyst (DOC), nitrogen monoxide (NO) is oxidized on the surface of noble metal catalysts. The function of a catalyst is to facilitate reactions by creating an easier path, thereby reducing the energy barrier required for the transformation. It is not consumed in the process and remains unchanged. However, under real operating conditions—particularly during NO oxidation—the catalyst surface undergoes nevertheless substantial structural and chemical modifications that directly impact NO conversion efficiency. This phenomenon, observed as "inverse hysteresis" in NO conversion, has been extensively studied using commercial catalytic converters containing Pt [1-6], Pd[5-7], and Pt–Pd alloy [8-10] catalysts. This behavior contrasts with the typical hysteresis observed in many oxidation reactions, such as CO and $C_3H_8$ oxidation. In typical hysteresis, during the heating phase, the reaction initiates at a higher ignition temperature, whereas during the cooling phase, the reaction persists until a lower extinction temperature is reached, resulting in higher conversion rates during cooling compared to heating. Such conventional hysteresis is often attributed to factors like thermal inertia, exothermic heat sustaining the reaction, and changes in the catalyst surface that maintain activity during cooling[1].

However, in the case of NO oxidation, an inverse hysteresis behavior indicates a different underlying mechanism. This phenomenon is likely attributable to the reversible oxidation and reduction of the platinum surface by $O_2$, $NO_2$, and NO, which are assumed to cause dynamic changes in surface coverage and catalytic activity during heating and cooling cycles. According to Rachel B. Getman et al.[9], the strength of the reductant critically affects the oxygen chemical potential and surface oxygen coverage: strong reductants such as CO and $H_2$ effectively remove surface oxygen, whereas weaker reductants like NO allow substantial oxygen coverage to remain.

At high temperatures, $O_2$ and $NO_2$ are believed to oxidize the platinum surface, decreasing its activity, while NO likely reduces the oxide back to metallic Pt at lower temperatures, restoring the activity. [1-6] [8-13]. Therefore, gaining insights into the surface transformations of noble metal catalysts is crucial for elucidating catalytic mechanisms and enhancing their practical performance.

Numerous experimental investigations have focused on the reaction of oxygen with Pt catalyst surfaces, employing advanced surface analytical methods such as low energy electron diffraction (LEED)[14-18], X-ray photoelectron spectroscopy (XPS)[14-17][19][20], high resolution electron energy loss spectroscopy (HREELS)[14], X-ray diffraction (XRD)[21,22], scanning tunnelling microscopy (STM)[19]. However, these analyses are typically performed on thin-film single crystal samples, which differ significantly from the geometry and operation conditions of applied catalysts, which are typical nanoparticles.

In addition to pure Pt, alloy catalysts — particularly binary or ternary systems based on platinum group metals (PGMs) — frequently experience substantial surface reconstruction and compositional degradation when subjected to severe oxidative conditions, which ultimately compromises their catalytic activity [20]–[22]. This challenge has led to the development of nano-engineered structures, such as core-shell alloys, to boost performance and reduce reliance on costly active metals. Consequently, achieving a fundamental atomic-scale understanding of these transformations is essential.

Several studies have further investigated surface transformations of Pt–Pd alloy catalysts under NO oxidation conditions. For example, M. Kaneeda et al. [23] employed X-ray diffraction (XRD) and transmission electron microscopy-energy dispersive X-ray spectroscopy (TEM-EDX) to show that, under NO oxidation at 300 °C, the addition of Pd systematically modified the chemical composition of the catalyst surface and effectively inhibited Pt sintering, thereby preserving catalytic activity. Besides that, J. Schütz et al. [24] used scanning transmission electron microscopy (STEM) to observe morphological transformations — such as the emergence of core–shell structures — and to measure particle growth during stepwise hydrothermal aging. EDX was then applied to map the spatial distribution and composition of Pt and Pd within the aged Pt–Pd/$Al_2O_3$ catalysts, linking these structural changes to the observed declines and eventual stabilization in NO oxidation activity. However, these studies focus primarily on how catalyst composition and particle size affect activity, and they lack precise experimental data on the oxidation state of the catalyst surface.

To accurately analyze the surface oxidation and structure of catalysts at the atomic scale, it is essential to employ analytical techniques that investigate samples that closely resemble the actual morphology of practical catalysts. Atom probe tomography (APT) [25]–[28] meets this requirement by enabling high-resolution, three-dimensional chemical imaging of nanometer-sized materials at near-atomic precision.

In APT, compositional analysis is achieved through time-of-flight mass spectrometry (TOF-MS). Atoms are field-evaporated from a needle-shaped specimen, typically triggered by laser pulses. As ions, they are accelerated by a high voltage and travel through a field-free region towards a position-sensitive detector. The mass-to-charge ratio of each ion is calculated based on its time of flight. At the same time, the lateral (x, y) impact positions of the ions on the detector allow determination of the spatial origin of each atom. By combining this positional information with the sequential order of evaporation (z-direction), a three-dimensional reconstruction of the specimen's atomic structure is generated. Advanced reconstruction algorithms may account for the tip's changing geometry, ion trajectories, and detection efficiency to generate accurate 3D atomic maps that reveal both chemical composition and spatial distribution at near-atomic resolution. As a result, APT delivers sub-nanometer spatial resolution and elemental sensitivity down to parts per million, making it a powerful tool for the nanoscale characterization of materials.

In the context of catalysis, APT offers a unique advantage: the analyzed sample mimics the morphology of an actual nanoparticle, typically exhibiting a hemispherical apex with a diameter of approximately 20 nm. This close resemblance enables atomic-scale analysis of catalyst surfaces under realistic geometries. Various studies have been recently performed to investigate how the surface of catalysts evolves in specific oxidative environments by using APT, such as for Pt-Rh, Pt-Ru [24]–[27] or pure Pd [7].

In this work, nanosized Pt and PtPd alloy samples were extensively analyzed during NO oxidation using a customized atom probe tomography (APT) setup, enabling a quantitative understanding of surface oxidation on the metal surface and their catalytic behavior. To replicate the conditions of the NO conversion, the same reaction atmosphere, temperature ramping profile, and target temperatures were reproduced in a reaction chamber directly integrated into the custom-built APT system [29][31]. This configuration enabled seamless sample transfer under ultra-high vacuum (UHV) conditions. Within this chamber, tip-shaped specimens composed of pure Pt and PtPd alloy were exposed to the reactive atmosphere of 500 ppm NO and 3% $O_2$ in $N_2$, heat-treated following the defined temperature program of the catalysis, and rapidly frozen after reaching 100 °C, 200 °C, and 300 °C during each respective operation cycle. Subsequent analyses were conducted using a laser-assisted APT instrument [32][33]. The resulting three-dimensional elemental distribution and the effective oxide thickness of the surface oxide layers were precisely characterized.

Furthermore, the oxide thicknesses determined by atom probe tomography (APT) are compared with the NO conversion ratios measured using a flat-bed reactor (FBR), in collaboration with the Institute of Chemical Process Engineering at the University of Stuttgart.

In the FBR setup, the catalyst is positioned in a metal block to ensure uniform heating and controlled gas flow for kinetic studies. The catalyst consisted of nano-sized particles synthesized via the spark discharge method [34]. The reactor temperature is maintained by individually regulated electric heating elements and rapidly cooled using compressed air. Prior to each measurement, the catalyst was pre-treated in 4% $H_2/N_2$ at 350 °C and subsequently tested for NO oxidation with a gas mixture of 3% $O_2$ and 500 ppm NO in $N_2$. The temperature was ramped from 80 to 450 °C at a rate of 3 K/min. The outlet gas composition was continuously monitored by FT-IR and mass spectrometry. For further details on the reactor configuration and procedure, see [5]. As a result, direct comparison of the oxide thickness with the NO conversion ratio provided clear insights into the oxidation dynamics and catalytic activity of Pt and PtPd catalysts.

## 2. Experimental Methods

### 2.1 Specimen Preparation for APT: From Wire to Needle-Shaped Tip

To enable field evaporation in atom probe tomography (APT), the sample must be shaped into a sharp, needle-like tip with an apex radius of less than 50 nm. This geometry concentrates the electric field at the tip apex, enabling the controlled removal of atoms from the surface during analysis. In this study, Pt and PtPd alloy tips are commonly fabricated from metal wires using techniques like electrochemical polishing and focused ion beam (FIB) milling, respectively.

Firstly, to prepare the Pt tips, a precise electrochemical polishing technique was utilized. The Pt wires (0.1 mm diameter, 99.9% purity, Chempur) were immersed in an electrolyte solution composed of 15% calcium chloride ($CaCl_2$) in a 1:1 mixture of water and acetone. A graphite rod served as the counter electrode. The procedure was conducted in two stages: an initial coarse polishing at 15 V AC to induce necking, followed by fine polishing at 5 V AC until the tip was fully formed. This method effectively sharpens the Pt wire to the desired apex radius of about 25 nm.

To prepare the PtPd alloy (0.1 mm diameter, 50 at% Pt–50 at% Pd, ESPI metals) for atom probe tomography (APT), annular milling was performed using a focused ion beam (FIB) system (Scios DualBeam, FEI)[35]. This technique involves concentric milling around the wire axis to form a sharp, needle-like tip. The process uses a gallium ion beam at 30 kV and 0.1 nA for initial shaping, the current is then reduced to achieve an apex radius of approximately 50 nm. A final polishing step at 5 kV and 48 pA is then applied to eliminate amorphous material from Ga implantation.

As shown in Figure1(left), the sample apex successfully mimics nanoparticles with a radius of approximately 20 nm and exhibits almost a hemispherical geometry. The atomic surface geometry was validated using Field Ion Microscopy (FIM), as exemplified in Figure 1 (right). The FIM image, acquired at 4.0 kV using He as the imaging gas, clearly displays the face-centered cubic (FCC) crystal structure. The observed pattern is indexed to show a (111)-orientation with a distinct 3-fold axis of rotational symmetry, confirming the suitability of the sample for APT analysis.

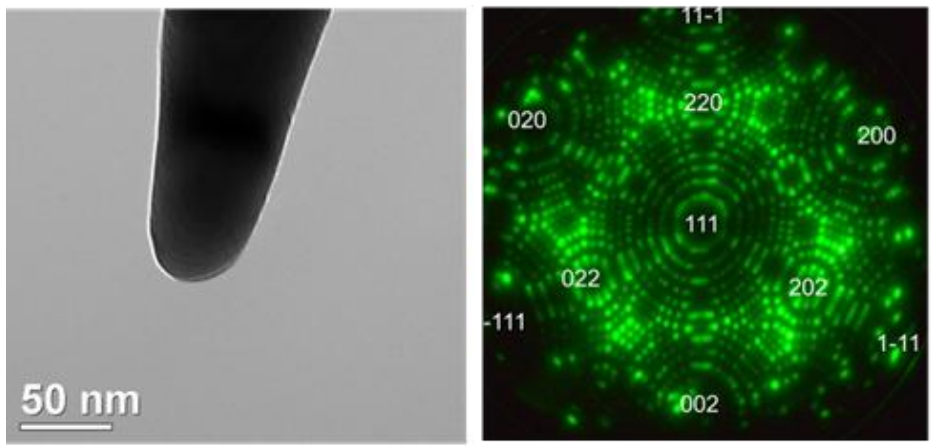

*Figure 1. TEM image of pure Pt tip fabricated by electrochemical polishing (left) and field ion microscopy of a rather sharp pure Pt tip at 4.0kV with He gas (right). Curvature radii for the shown tips amount to about 25 nm and 12 nm, respectively*

By evaluating the ring count between lattices poles of known orientation, the curvature of the surface can be precisely evaluated. In the shown case of Fig. 1 (right), the surface is slightly elliptic with curvature radius varying between 10 and 12 nm.

## 2.2 Gas reaction and APT measurement

This study used a custom-made APT instrument at the IMW (Institute of Material Science in the Stuttgart University) [32][33]. Aiming for a better time resolution and accuracy for light elements, the instrument is equipped with an uncommonly long flight path (305 mm). In consequence the opening angle of the effective aperture is limited to about ±15°. To allow the study of surface reactions the instrument is equipped with a dedicated reaction chamber for surface treatments.

To accurately measure the change of surface states of noble metal catalysts, a system enabling gas-phase reactions is necessary. Furthermore, guaranteeing sample transfer to the APT instrument under ultra-high vacuum (UHV) is critical to avoid surface contamination. Therefore, an oven chamber dedicated to oxidation reactions was directly integrated, maintaining UHV conditions ($p < 10^{-8}$ mbar) throughout the process [7]. Additionally, various gases and control systems were installed to effectively manage operation conditions similar to the experiments in a flat-bed reactor. The integrated setup ensures that the entire workflow—from initial surface cleaning via field evaporation to gas exposure and subsequent APT analysis—occurs within a single instrument, preserving sample integrity and providing reliable data on surface and subsurface oxidation states.

To ensure accurate analysis of surface processes, it is essential to eliminate contaminants initially present on the tip surface. Therefore, prior to the oxidation cycles, the tips underwent preliminary field evaporation within the APT. This step effectively removed residual impurities, particularly those introduced during sample preparation, as evidenced by the initial mass spectra. Once a clean surface was confirmed in the APT mass spectra, the samples were transferred to the gas reaction chamber without breaking the UHV conditions, thereby preserving their pristine state for the catalysis.

In the reaction chamber, samples experienced a temperature ramp under an atmosphere of 0.05% NO and 3% $O_2$ in $N_2$. Figure 2(i) depicts the linear temperature ramps, operating between 80 ℃ and 450 ℃ with 3 K/min rate. Samples were intentionally quenched at 100 °C, 200 °C, and 300 °C during the first heating (1a), first cooling (1b), and second heating (2a) ramps. These quenching points are distinctly mapped using specific symbols (circle (○), diamond (◇) and triangle (△)) and colors (yellow, orange, red) corresponding to the ramp and temperature.

At each stage, rapid cooling was applied to freeze the samples while simultaneously evacuating residual gases, as shown in Figure 2(ii) for the example of a sample quenched from 300 °C during the first cooling cycle. Clearly proven differences between various quenched states indicate that the quenching rate is sufficient to conserve the desired reactions states for measurement. Following the oxidation, the samples were reinserted into the measurement chamber for APT analysis maintaining ultra-high vacuum (UHV) during transfer which took about 2 min.

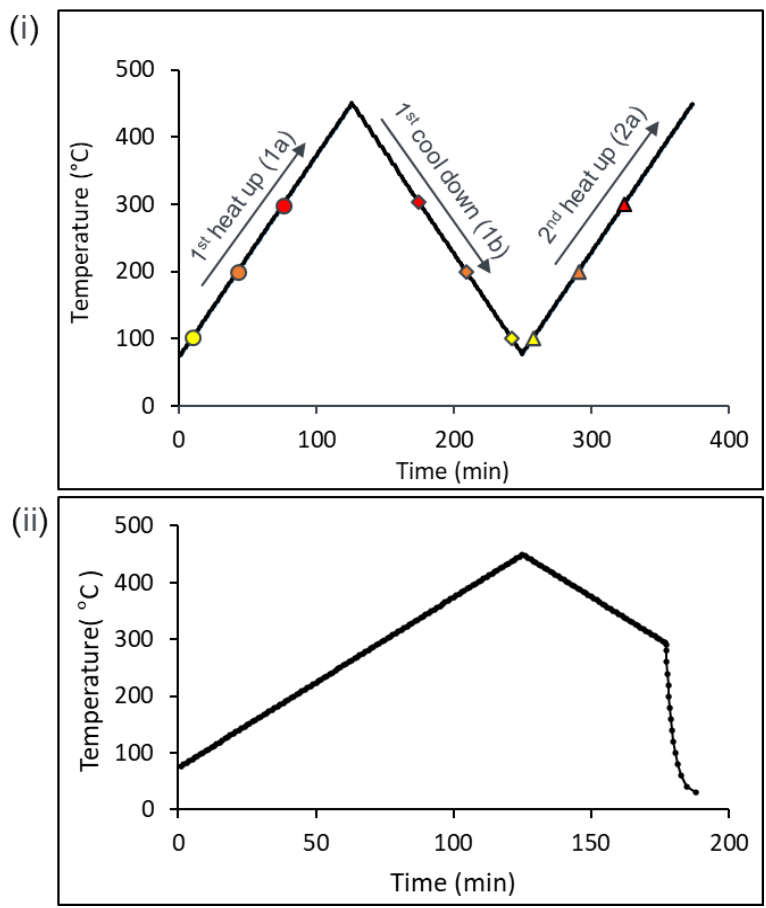

*Figure 2. (i) Scheme of temperature ramp for NO conversion measurement of a Pt sample. The samples were intentionally quenched at three specific temperatures, i.e.100 °C (yellow), 200 °C (orange), and 300 °C (red), during three distinct thermal cycles: the first heating (1a), the first cooling (1b), and the second heating (2a) ramps. These quench points are clearly mapped in the data using specific symbols: circle (○), diamond (◇) and triangle (△). (ii) Heating and cooling cycle for the sample annealed to 1st cooling at 300 ℃. Quenching (dotted line) rate is fast to freeze the samples. The experiment was conducted in a gas environment of 500 ppm NO and 3% $O_2$ in $N_2$.*

Within the measurement chamber, the sample was cryogenically cooled to 38 K under ultra-high vacuum environment ($10^{-10}$ mbar), thereby effectively immobilizing the specific surface atomic arrangement. APT measurement was conducted using an ultrafast UV laser system (Clark-MXR), which generated 250 fs pulses [32]. The frequency of the laser was set to 100 kHz at an average power of 15 mW equal to 150 nJ pulse energy. To ensure reproducibility, APT analyses were independently repeated with freshly prepared surfaces, 3 to 5 times at each recorded stopping point across the different temperature ramps.

The measured volumes were reconstructed using the SCITO software package[36]. Reconstruction employed a conventional geometrical algorithm [37] based on Bas et al.'s original point projection method (1995).

This study aims to compare catalyst surface changes during NO conversion. For this purpose, degree of oxidation was quantified defining an *effective* oxide thickness ($\mathcal{T}_{\text{eff}}$)

$$\mathcal{T}_{eff} = \frac{n_O}{A \times \rho \times \mathcal{D}} \, [nm] \qquad \qquad \textit{Equation 2-1}$$

where, $n_O$ denotes the number of oxygen atoms, determined by integrating the oxygen-containing peaks in the mass-to-charge histogram, $A$ is the area of the apex of reconstructed volume, $\rho$ is the atomic density of oxygen within the oxide and $\mathcal{D}$ is the calibrated detector efficiency of the microchannel plate used in the APT ($\mathcal{D}$ = 0.43). The stronghold of APT is the analysis of local stoichiometry. Since the structure cannot be discovered and the atomic density varies among different oxides, we decided to use here always the specific oxygen density of PtO to warrant comparability between different samples and reaction stages. In this sense, the effective oxide thickness has to be seen primarily as a measure of the absolute oxygen coverage, which can however exceed a single monolayer.

## 3. Results and discussion

### 3.1 Effective oxide thickness of pure Pt

The mass spectra obtained from the pure Pt samples subjected to the initial heating (1a) to 200°C (1a_200°C), further heating to 450 °C and subsequent cooling down to 200°C (1b_200°C), and second heating to 100°C (2a_100°C) in a gas mixture of 500 ppm NO and 3% $O_2$ in $N_2$ are depicted in Figure 3 (i), (ii), and (iii), respectively. The mass spectra recorded during the initial heating phase (1a) up to 200°C, exclusively identify oxygen ions ($O^+$ and $O_2^+$) alongside platinum ions ($Pt^+$, $Pt^{2+}$). A major transformation is evident once the temperature surpassed 300°C, where the spectra showed the emergence of new species: platinum-oxygen molecules ($PtO^+$, $PtO_2^+$). After heating to 450°C, the system initiated the cooling ramp(1b). Samples subsequently quenched at 300°C, 200°C, and 100°C revealed further complexity. For instance, the spectrum of the sample cooled to 200°C as shown in Figure 3(ii) also contains oxygen cluster ions ($^{48}O_3^+$, $^{64}O_4^+$ and $^{80}O_5^+$). Conversely, the mass spectra acquired at the start of the second heating phase (2a) following complete cooling, as shown in Figure 3(iii), indicate a clear diminution in oxygen-related signals.

Table 1 summarizes the quantitative data on the effective oxide thickness. During the first heating(1a) to 200°C, oxide growth is minimal. However, heating to 300°C, triggers substantial oxidation. After the peak temperature of 450°C and cooling back to 300°C, the oxide thickness has increased to 0.52 ± 0.01 nm. This growth trend is sustained throughout the first cooling ramp (1b) until the temperature reaches 200°C.

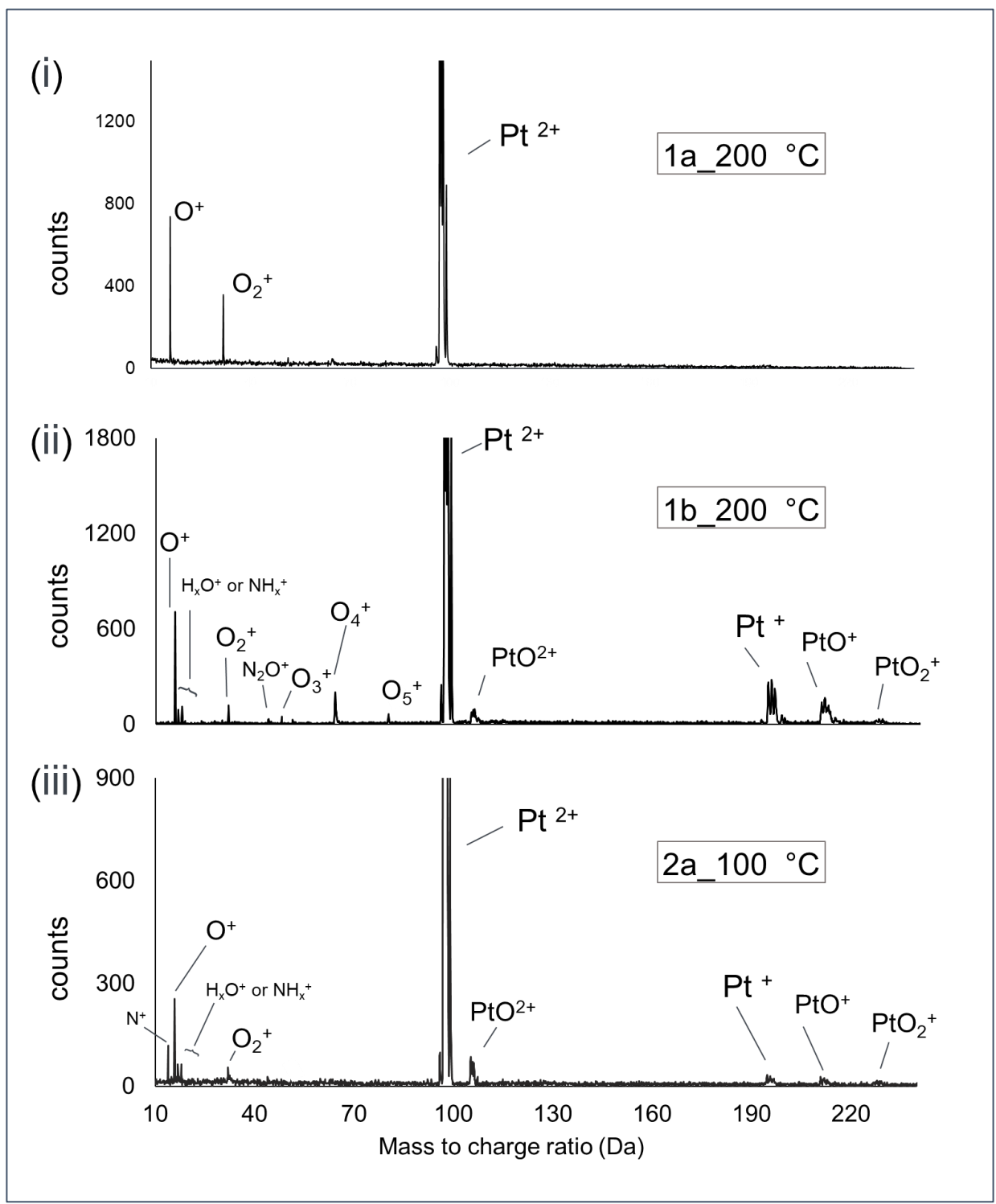

*Figure 3. Mass spectra obtained in the initial heating ramp after reaching 200°C (i), subsequent cooling to 200°C (ii), and 2nd heating to 100°C (iii) in a gas mixture of 500 ppm NO and 3% $O_2$ in $N_2$ of pure Pt samples.*

Table 1. Effective oxide thicknesses of pure Pt as determined by APT after freezing the samples at 100°C, 200°C, and 300°C during a temperature ramp from 80°C to 450°C at 3 K/min. (Error margins were determined from comparison of repetition measurements with independent samples)

| Cycle | Temp(°C) | Oxide thickness (nm) |
|---|---|---|
| 1a (Heating) | 100 | 0.14 ± 0.01 |
|  | 200 | 0.09 ± 0.01 |
|  | 300 | 0.24 ± 0.02 |
| 1b (cooling) | 300 | 0.52 ± 0.01 |
|  | 200 | 0.57 ± 0.03 |
|  | 100 | 0.47 ± 0.02 |
| 2a (Heating) | 100 | 0.35 ± 0.01 |
|  | 200 | 0.40 ± 0.01 |
|  | 300 | 0.49 ± 0.01 |

An interesting feature regarding the oxygen distribution on the catalyst surface was observed. Figure 4 presents the 3D volume reconstructions and compositional analysis for selected cylindrical sub-volumes of the Pt sample after the first cooling down to 300 °C (1b_300°C). In particular, a comparison between oxygen-rich and oxygen-poor regions is highlighted. Although the oxide layer is less than 1 nm thick, the distribution of oxygen on the catalyst surface shows significant variations.

APT data naturally exhibit patterns similar to stereographic projections in FIM, reflecting variations in atomic density across crystallographic planes and zone axes. Such features could reveal facet-dependent oxide formation. However, the used atom probe has an aperture angle of only ±15°. Thus, the scanned surface area represents just the extended top facet of the central (111) pole and at most a small number of nearby (111) terraces. Since the chemical concentration does not show any correlation with the distance to the pole, the observed variation reveals random fluctuations as expected from a homogeneous nucleation process on the surface. To quantify these fluctuations, a statistical analysis on the local oxygen content in circular regions of 5 nm diameter was conducted on the surface data (the first 18% of the complete data set primarily corresponding to the oxide region while excluding the pure Pt underneath). The bottom graph in Figure 4 presents the resulting local concentration statistics for oxygen in comparison to a binomial distribution of a random atomic arrangement. Clearly, the experimental histogram exhibits a noticeably broader distribution than a binomial. Furthermore, a visible shoulder indicates possibly a bimodal concentration distribution. This suggests real nanoscale clustering in the oxide region, rather than purely statistical fluctuations. Possibly, nucleation of a second (surface) oxide phase is proceeding here.

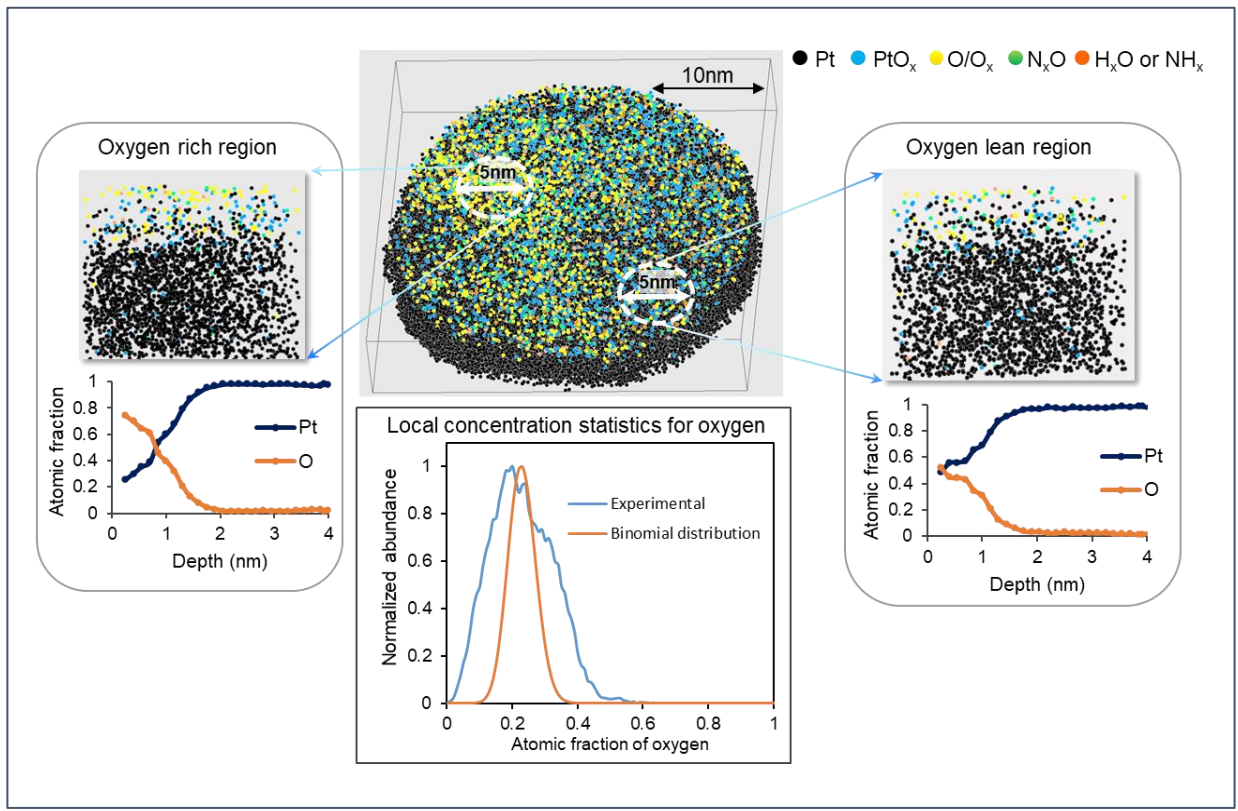

*Figure 4. 3D Volume reconstructions and composition profiles for selected cylindrical sub-volumes for the Pt sample after the first cooling down to 300 °C (1b_300°C) in a 500 ppm NO and 3% $O_2$ in $N_2$ environment, highlighting a comparison between oxygen-rich and oxygen-poor regions. Bottom diagrams shows the distribution of concentrations in disks of 5 nm diameter.*

Further cooling to 100°C reveals a minor decrease in the oxide thickness, which becomes, however, more pronounced during further cooling as the initial state of the second heating ramp (2a) demonstrates even thinner oxide. This suggests that significant oxygen reduction occurs below 100 °C. Conversely, as the temperature increased above 100°C during the second heating (2a), the oxide layer begins to regrow. This regrowth ultimately resulted in an oxide layer even thicker than that observed during the initial heating phase (1a).

## 3.2 Correlation between Oxide Thickness and NO Conversion on pure Pt

The effective oxide thickness measured by Atom Probe Tomography (APT) is directly compared with the NO conversion ratio obtained from a Fixed-Bed Reactor (FBR) coupled with FT-IR spectroscopy. The reactor was loaded with active Pt particles of similar curvature radius than the atom probe tips.

Figure 5(i) illustrates the pure Pt performance. During the initial heating ramp (1a), the NO conversion remains minimal, showing little activity below ~150°C. Then, the conversion rate rises between 150°C and 350°C and finally it decreases beyond 350 °C, which is attributed to the thermodynamic constraint of the $NO + 1/2\ O_2 \leftrightarrow NO_2$ reaction [3]. Interestingly, the subsequent cooling ramp (1b) exhibits a significant drop in conversion efficiency at equivalent temperatures, signifying a considerable decline in catalytic function.

This observed change in kinetics aligns strikingly with the measured effective oxide thickness, as presented in Figure 5 (ii). Initially, the surface oxide layer is negligible, with a thickness of around 0.1 nm, suggesting that the catalyst surface was largely metallic. However, oxide formation sets in at 200°C during heating and continues to grow until the temperature reaches 300°C. When the sample reached the 450°C peak and transitioned into the cooling phase (1b), the oxide layer had grown substantially, exceeding 0.5 nm. Since the emergence of this oxide layer closely correlates with the observed reduction in NO conversion, it is strongly suggested that the oxide-covered surface evidently limits the material's catalytic efficiency.

Probably, the partial recovery of NO conversion rate is the most significant observation. As depicted in Figure 5(i), a marked rise in NO conversion is observed beyond 150°C in the second heating ramp, confirming that some catalytic activity has been obviously restored during the previous cooling stage. The measured oxide thickness data clearly elucidates this behavior. Indeed, following cooling (1b) to a minimum of 80°C and subsequently reheating (2a), a partial reduction in the surface's oxidation state is confirmed. Clearly, the oxide layer does not vanish completely, yet its discernible thinning directly supports the partial recovery of NO conversion observed in the second cycle.

Further support for this correlation comes from comparing NO conversion and oxide thickness at 300°C. The NO conversion rates are similar between the first cooling ramp (1b) and the second heating ramp (2a), and so are the oxide thickness values, as seen in Figure 5(ii), which also indicated a comparable oxidation state at 300°C across both cycles.

As a further confirmation, figure 5(iii) presents for comparison the theoretical predictions for the metallic Pt surface fraction during the NO conversion ramp, as derived from a rate kinetic models that were developed by H. Dubbe et al. in 2016 [5]. It shows the predicted remaining noble metal fraction on a Pt catalyst exposed to a gas mixture of 500 ppm NO, 5% $H_2O$, 7% $CO_2$, and 5% $O_2$ in $N_2$. First heating(1a) starts with 100% metallic platinum. The model predicts that the metallic Pt fraction remains stable up to 150°C during the initial heating ramp (1a), then sharply declines when heating up to 350°C. But remarkably, above 350°C, the Pt fraction recovers partly. Throughout the cooling phase (1b), the Pt fraction remains very low, indicating a predominantly oxidized surface. Below 150°C (both cooling and heating), a recovery in the metallic Pt fraction occurs, corresponding to the partial recovery of catalytic activity. However, the fraction of metallic Pt becomes very low again upon reaching 300°C, suggesting a reversible oxidation-reduction cycle. Obviously, the rate kinetics model's predictions for the metallic Pt fraction are highly consistent with the measurements of oxide thickness, although from a slightly different point of view. Unlike the theoretical work, which considered coverage by a single monolayer of oxygen, atom probe data clearly demonstrate the presence of a somewhat thicker (approximately 0.6 nm, two to three lattice planes) oxygen rich layer or oxide phase that covers the surface, and its thickness controls the conversion rate.

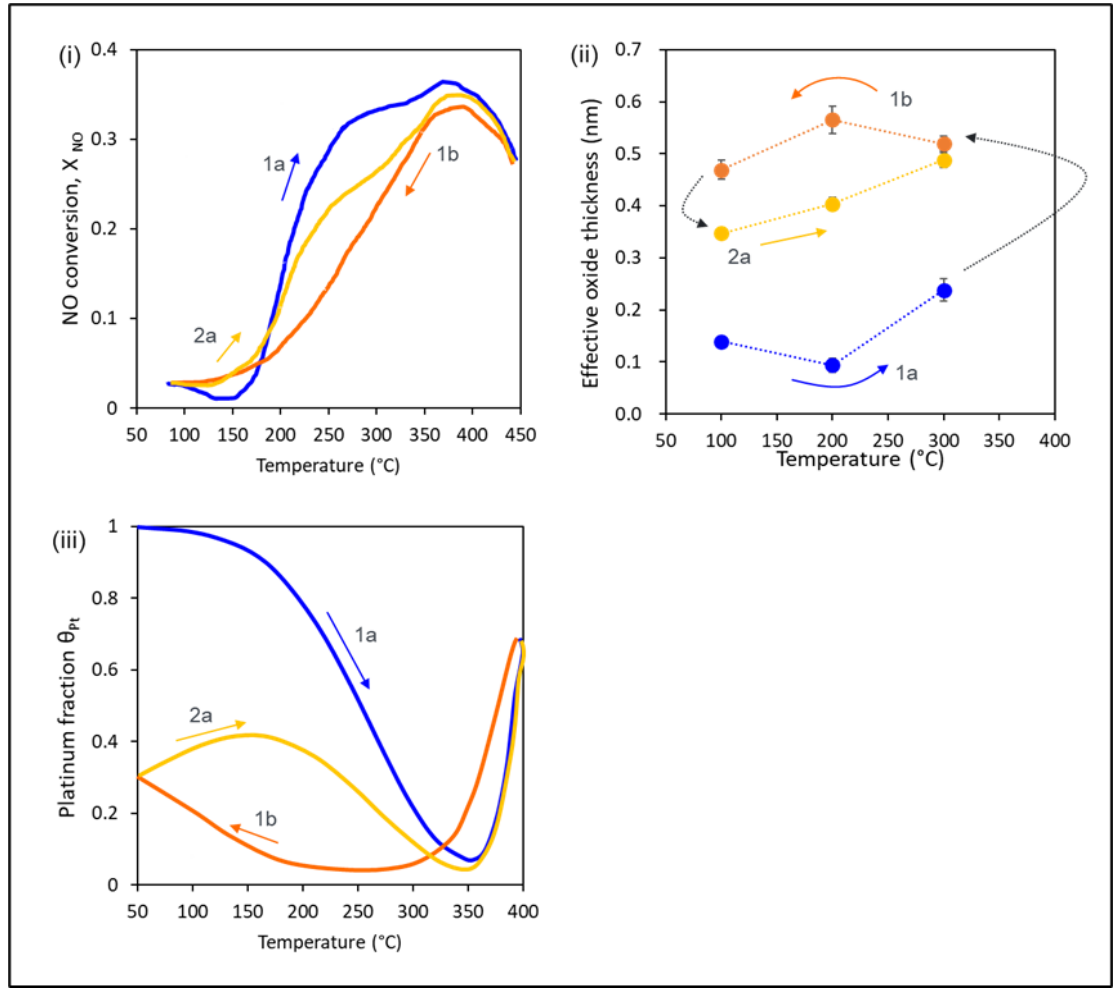

*Figure 5. Diagram of (i) the NO conversion rate as function of temperature measured by FBR under a gas atmosphere of 500 ppm NO and 3% $O_2$ in $N_2$, (ii) the effective oxide thickness of Pt samples determined using APT at various quenched temperatures. Samples were subjected to the same gas atmosphere as in the FBR measurement (500 ppm NO and 3% $O_2$ in $N_2$), (iii) Simulation results of the noble metal fraction of Pt on the Pt-catalyst in a gas mixture of 500 ppm NO, 5% $H_2O$, 7% $CO_2$, and 5% $O_2$ in $N_2$ done by H. Dubbe et al.[5].*

## 3.3 Effective oxide thickness of 50at%PtPd alloy

As with pure Pt, the 50 at% PtPd alloy samples exhibited similar dynamic changes in their oxidation state during the temperature ramp under the same gas mixture (500 ppm NO and 3% $O_2$ in $N_2$). Figure 6 shows for comparison representative mass spectra of a 50at%PtPd alloy tip subjected to three sequential thermal treatments in the same gas mixture containing 500 ppm NO and 3% $O_2$ in $N_2$: (i) the initial heating phase to 200°C (1a_200°C), (ii) the subsequent cooling phase to 200°C (1b_200°C), and (iii) the second heating phase to 100°C (2a_100°C). The first heating cycle to 200°C demonstrates mass spectra where the prevailing signals were attributed to complex oxygen species. Unusual, ionic clusters such as $O_2^+$ (32 Da), $O_4^+$ (64 Da), and $O_8^+$ (128 Da) were detected.

After reaching a maximum temperature of 450°C, followed by cooling down to 200 °C, mass spectra are obtained as displayed in Figure 6(ii). They clearly reveal the formation of platinum oxide species, such as $PtO^+$ and $PtO_2^+$, while in contrast no distinct palladium oxide peaks are detected. The most distinct feature is the consistent presence of oxygen cluster ions, specifically $^{64}O_4^+$, and $^{128}O_8^+$. These striking species are observed without exception across all subsequent temperature ramps, including the first cooling and second heating phases, except for the sample that underwent the temperature ramp up to 2a_100°C. The representative mass spectrum from this phase (2a_100 °C), shown in Figure 6(iii), highlights significant changes in the surface chemistry. Notably, the previously observed oxygen cluster ions disappeared, while signals corresponding to atomic oxygen and platinum oxide species become prominent.

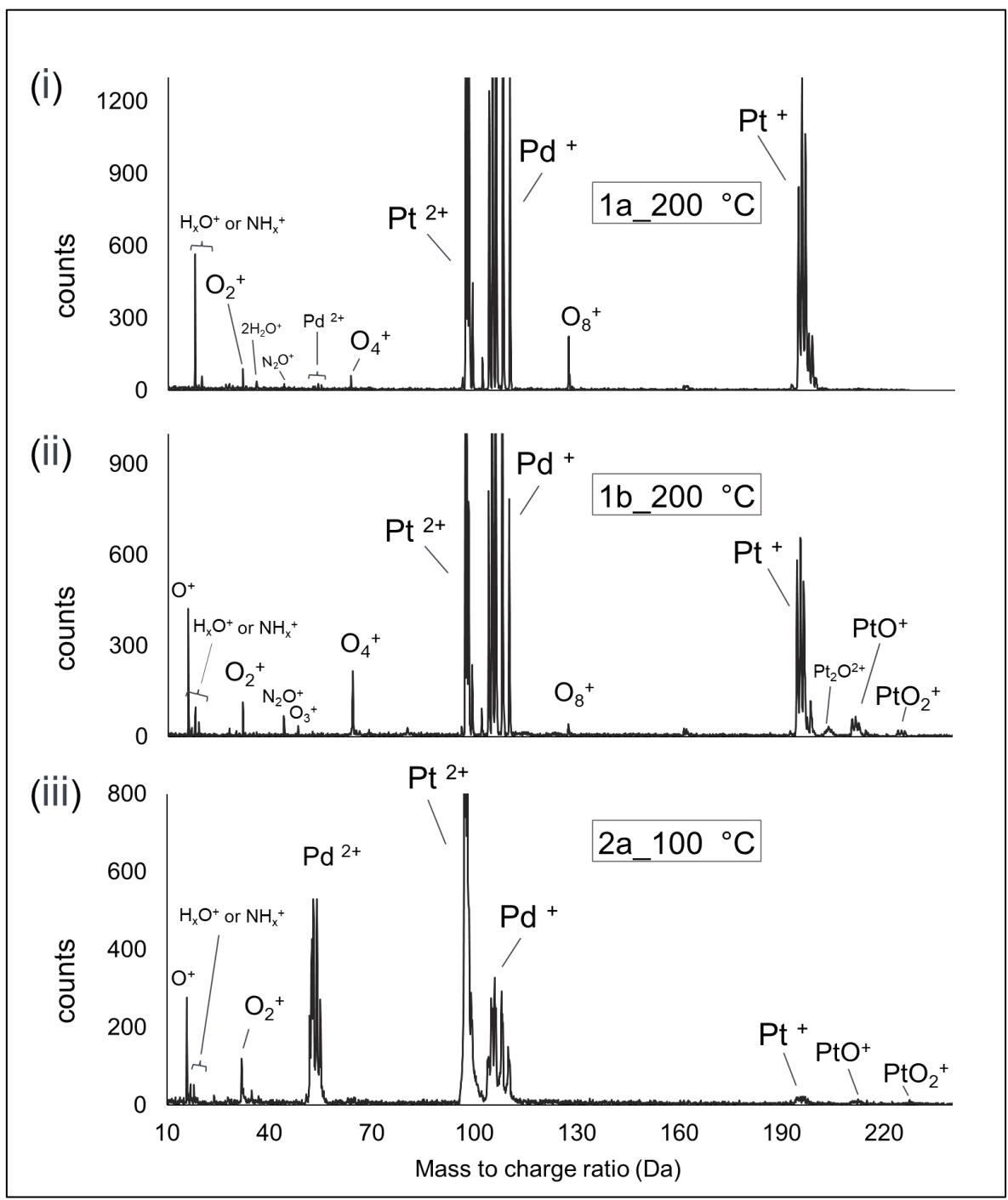

*Figure 6. Mass spectra obtained in the initial heating ramp after reaching 200°C (i), after heating to 450°C and subsequent cooling to 200°C (ii), and 2nd heating to 100°C (iii) in a gas mixture of 500 ppm NO and 3% $O_2$ in $N_2$ of 50at%PtPd alloy samples. (Note: the striking variation of the $Pd^{2+}:Pd^{1+}$ ratio between the states is a consequence of the laser measurement conditions. It is not strictly linked to the sample stage.)*

The effective oxide thicknesses for PtPd alloy samples are presented in Table 2. Oxide formation is minimal during the initial heating ramp (1a) up to 100°C. A significant oxidation occurs as the temperature rises from 200 to 300 °C. Following subsequent heating to 450°C and then cooling back to 300°C, the oxide thickness increases notably to significant 1.10 ± 0.02 nm, almost a factor 2 thicker than with pure Pt. This oxidation growth continued even during the early stages of the first cooling ramp (1b) down to 200°C.

Table 2. The effective oxide thicknesses on 50at% PtPd alloy determined using APT after freezing the samples at 100°C, 200°C, and 300°C during a temperature ramp from 80°C to 450°C at 3 K/min.

| Cycle | Temp(°C) | Oxide thickness (nm) |
|---|---|---|
| 1a (Heating) | 100 | 0.15 ± 0.01 |
|  | 200 | 0.42 ± 0.03 |
|  | 300 | 0.77 ± 0.01 |
| 1b (cooling) | 300 | 1.10 ± 0.02 |
|  | 200 | 1.11 ± 0.03 |
|  | 100 | 1.04 ± 0.01 |
| 2a (Heating) | 100 | 0.59 ± 0.01 |
|  | 200 | 0.70 ± 0.02 |
|  | 300 | 0.80 ± 0.02 |

Further cooling down to 100 °C results in a slight decrease in the oxide layer thickness, which has become more distinct (effective oxide thickness of 0.59 nm) at the start of the second heating phase (2a). This confirms a partial reversal of the oxidation. However, with subsequent further rise of temperature in the second heating cycle, the oxide layer regrew, eventually even exceeding the thickness established during the initial heating phase (1a).

The most intriguing finding at the PtPd alloy is the reorganization of Pd atoms, specifically their segregation to the surface. To track this surface compositional change, composition profiles were derived from the Atom Probe Tomography (APT) 3D reconstructions, as detailed in Figure 7. The figure showcases three distinct snapshots:

the sample state at 200°C during initial heating (1a), the state at 200°C during cooling (1b), and the final state after reheating 100°C (2a).

Figure 7(i) shows that up to 200°C in the initial heating phase, the Pt to Pd atomic ratio remained near 50:50, despite an increase in surface oxygen.

Figure 7(ii) presents a striking contrast: Pd atoms diffused to the surface, establishing an Pd-O rich layer approximately 1nm thick, following the 450°C heat treatment and subsequent cool-down to 200°C. This Pd enrichment led to a corresponding, though difficult-to-quantify Pd depletion zone located beneath the rich layer.

This surface segregation establishes during the first heating ramp once the temperature passed 300°C. Figure 7(iii) depicts the surface evolution during the second heating process (2a), revealing two notable features as the sample is reheated to 100°C: First, as confirmed by both the composition profile and 3D reconstruction images, there is a significant reduction in the surface oxygen concentration. But second, Pd segregation to the surface remains evident. Clearly, although the oxidation is largely reduced, the segregation of Pd to a composition of about $Pd_{66}Pt_{33}$ remained.

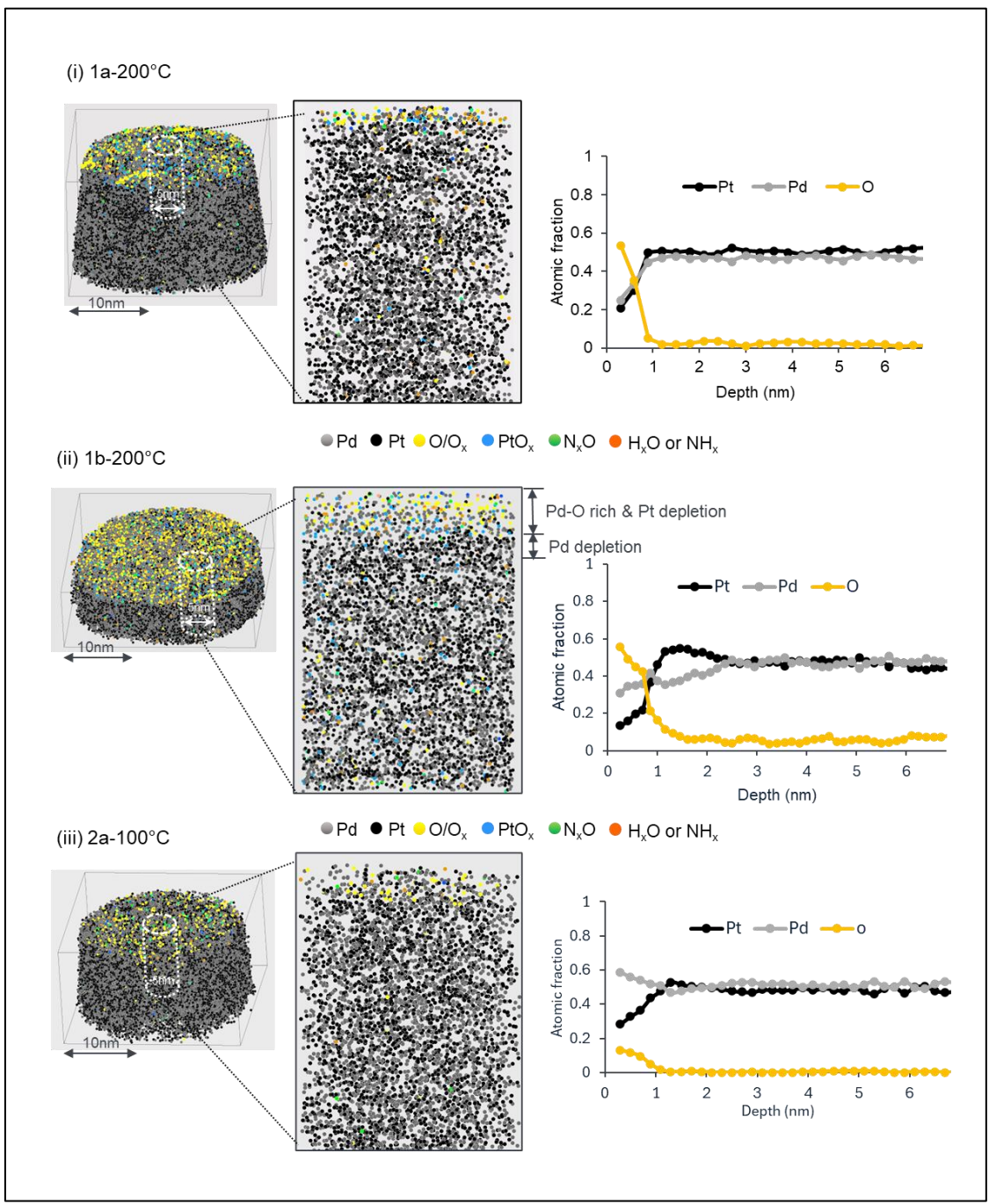

*Figure 7. 3D volume reconstructions and depth composition profiles for selected cylindrical sub-volumes. (i) for the sample heated to 200°C in the heating phase (1a), (ii) for the sample cooled to 200°C in the cooling phase (1b), and (iii) for the sample heated to 100°C in the 2$^{nd}$ heating phase (2a). Note, due to the limited aperture area of an atom probe, the analysis does not comprise the full tip, but just a central cone. Thus, only a spherical cap of the surface measured over a full polar angle of 30° is comprised, while the flank sides of the atomic reconstructions represent inner volume of the tip. Consequently, no oxidation is observed there.*

The observation of palladium migrating to the surface at the relatively moderate temperatures of 300°C to 450°C represents a pivotal discovery in surface catalysis. It is particularly notable because the PtPd binary system is recognized for its sluggish diffusion kinetics; its diffusion coefficients at 400 °C typically range from $10^{-25}$ to $10^{-24}$ $m^2s^{-1}$ [38][39][40]. For example, assuming a diffusion coefficient of $10^{-25}$ m²/s at 400 °C and a heat treatment duration of 80 minutes, the estimated diffusion length is 0.098 nm. This value is much lower than e.g. observed in systems like Pt–Fe, which reach diffusion coefficients on the order of ~$10^{-24}$ $m^2s^{-1}$ even at 300 °C [41]. However, the Pd diffusion is clearly promoted under the elevated driving force of the oxidizing environment used in the present study.

The observed Pd segregation aligns with findings by Tong Li et al. in 2012 [20], who investigated a Pt (69%) - Pd (31%) alloy. Under significantly harsher oxidative conditions (1 bar O2 at 400°C to 800°C for 5 hours), they noted the formation of a Pd-rich chemisorbed oxygen layer on the tip surface, particularly in the similar temperature range of 400°C to 500°C.

Supporting simulation studies [40] [41] also confirmed the preferential segregation of Pd toward the surface of Pt–Pd alloys. This tendency originates mainly from Pd's lower surface energy (1.9 J/m²) compared with that of Pt (2.6 J/m²) for nanoparticles of approximately 1.9 nm in diameter [42]. Nevertheless, the theoretical prediction of elemental segregation at the surface is not straightforward. As highlighted by the density-functional theory (DFT) calculations of Arezoo et al.[44], competing effects may gover segregation behavior. Under vacuum, Pt's higher electronegativity (Pt: 2.28 vs. Pd: 2.20 on the Pauling scale) drives it toward the outer surface due to d-band filling effects, while its greater surface energy concurrently favors migration into the subsurface region. Thus, in the absence of oxygen, the surface stability is composition-dependent: Pt stabilizes the surface in Pt-rich alloys, whereas Pd becomes dominant on the surface in Pd-rich alloys.

Crucially, the DFT results indicate that an oxygen ad-layer is essential to promote Pd segregation. This outcome stems from complex electronic interactions and shifts in the d-band center, which modulate oxygen adsorption strength. When the less electronegative Pd occupies the surface and Pt resides beneath, the surface d-band filling decreases, leading to enhanced oxygen binding and elevated surface energy— an effect that aligns well with the d-band center theory [45].

Beside subtle electronic effects, thermodynamic parameters reinforce the fundamental cause of oxidation induced or accelerated segregation. The Gibbs free energy of formation for PdO is more negative than that for PtO, as shown in the Ellingham diagram [46][47], making the oxidation of palladium thermodynamically favorable. The difference in stability is reflected in the heats of formation reported in the literature as -142.26 kJ/mol for PtO [48] and -175.72 kJ/mol for PdO [20] for one mole of oxygen.

Consequently, in an oxygen-rich atmosphere, Pd is driven toward the surface because its strong thermodynamic tendency to form stable oxides (favorable Gibbs energy) is amplified by electronic structure changes (d-band center shifts) that facilitate robust oxygen binding. Quantitative validation of this behavior, summarized in Table 3, was derived from the outermost surface atomic fractions of Pt and Pd (calculated both with and without oxygen) in the composition profiles (Figure 7). Specifically, the amount of Pd nearly doubled that of Pt after the 1b_200°C stage, confirming clear Pd surface segregation.

Table 3. Quantified surface atomic fractions of Pt and Pd for samples subjected to different temperature ramps: first heating to 200 °C(1a_200°C), first cooling to 200 °C(1b_200°C), and second heating to 100 °C(2a_100°C)

|  | With Oxygen | | | Without Oxygen | |
| --- | --- | --- | --- | --- | --- |
|  | Pt | Pd | O | Pt | Pd |
| 1a_200 °C | 0.23 ± 0.02 | 0.26 ± 0.02 | 0.51 ± 0.04 | 0.48 ± 0.01 | 0.52 ± 0.01 |
| 1b_200 °C | 0.16 ± 0.01 | 0.35 ± 0.004 | 0.49 ± 0.01 | 0.31 ± 0.01 | 0.69 ± 0.01 |
| 2a_100 °C | 0.28 ± 0.01 | 0.58 ± 0.004 | 0.14 ± 0.004 | 0.33 ± 0.01 | 0.67 ± 0.01 |

Similar to the pure Pt surfaces, the three-dimensional reconstruction in Figure 7(i) reveals a clear spatial variation in the initial oxygen distribution over the PtPd alloy surface, suggesting that oxide growth proceeds in a locally heterogeneous manner rather than uniformly.

To provide a quantitative description of this behavior, localized compositional analyses of the oxide layer were carried out for three distinct experimental states corresponding to Figure 7: (i) the specimen heated to 200 °C during the first heating step (1a), (ii) the specimen cooled to 200 °C in the subsequent cooling stage (1b), and (iii) the specimen reheated to 100 °C in the second heating cycle (2a). For each condition, the surface portions of the datasets—namely 4–20%, 7–23%, and 6–13%, respectively—were extracted with emphasis on the oxide region, and the statistical distributions are compared with a binomial model. As illustrated in Figure 8(i), the obtained abundance histograms exhibit markedly broader profiles than those predicted by the binomial expectation, implying strong local fluctuations in oxygen concentration and indicating that oxidation proceeds through a non-random mechanism at the nanoscale.

In contrast with prior observations on pure Pd[29], where oxygen was reported to be homogeneously distributed, the PtPd alloys display behavior more comparable to that of pure Pt. With continued temperature cycling through the cooling and second heating stages, however, the deviation from the binomial trend becomes progressively smaller, which is likely a consequence of increased Pd segregation toward the surface.

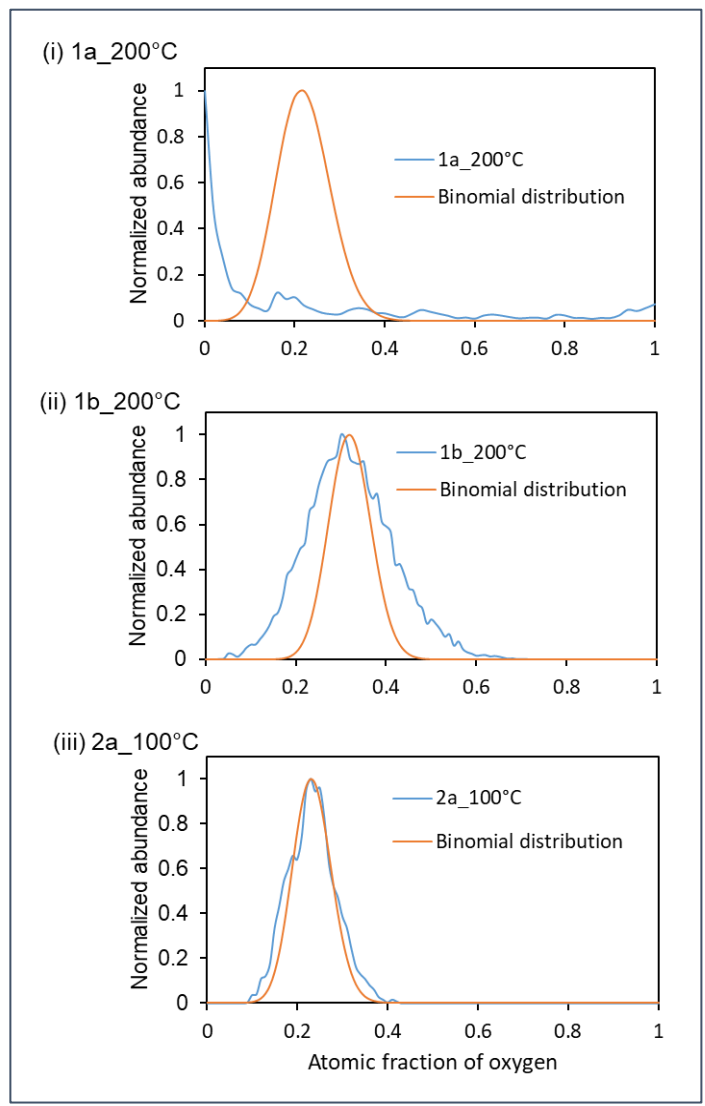

*Figure 8. Local concentration analysis of oxide layer (i) for the sample heated to 200°C in the heating phase (1a), (ii) for the sample cooled to 200°C in the cooling phase (1b), and (iii) for the sample heated to 100°C in the 2<sup>nd</sup> heating phase (2a). For each thermal condition (1a, 1b, and 2a), the statistical distribution of the oxide was determined by selectively analyzing the early data, namely 4–20%, 7–23%, and 6–13%, respectively, and the statistical distributions were compared with a theoretical binomial model*

## 3.4 Correlation between Oxide Thickness and NO Conversion on PtPd alloy

The redox behavior of the PtPd alloy in a reactive gas mixture containing 0.05% NO and 3% $O_2$ balanced with $N_2$ exhibit a temperature-dependent evolution of the oxidation state, closely resembling that observed for pure Pt. This temperature-driven variation also corresponds well with the NO conversion profiles, confirming a direct link between oxidation dynamics and catalytic performance. As illustrated in Figure 9(ii), during the first heating ramp (1a), the conversion of NO remains minimal until approximately 140°C, beyond which a pronounced rise occurs, signifying catalyst activation. Subsequently, in the first cooling phase (1b), a marked decline in conversion efficiency is detected. According to Figure 9(i), this reduction in activity coincides with the formation of an oxide overlayer, which probably inhibits surface reactivity. The oxide, reaching a thickness of about 1.1 nm, is noticeably thicker than that observed on a pure Pt catalyst.

During the second heating sequence (2a), the NO conversion rate shows partial recovery, qualitatively mirroring the trend found for pure Pt. This recovery, occurs mainly below 100 °C between the cooling (1b) and subsequent heating (2a) steps. The evolution of oxide thickness provides a coherent explanation for this phenomenon. As the temperature rises above 100 °C, the oxide gradually thickens, attaining a maximum of roughly 0.77 nm at 300 °C. Upon heating to 450 °C and subsequent cooling (1b), the oxide continues to grow, exceeding 1 nm, and remains nearly unchanged down to 200 °C. Below this temperature, however, the effective oxide thickness begins to shrink.

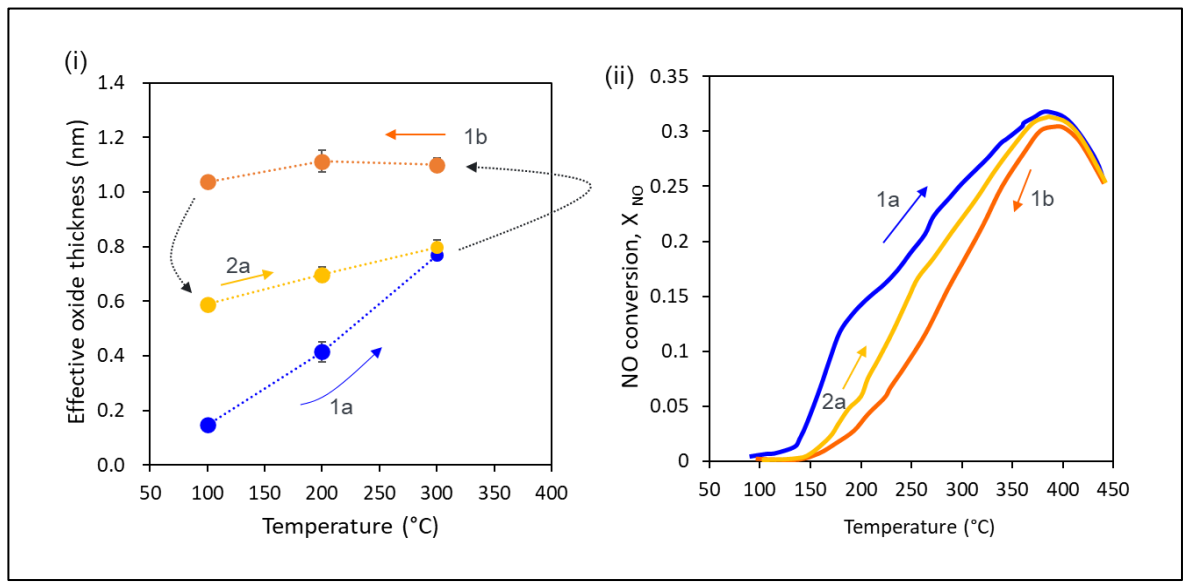

*Figure 9. Diagram of (i) the effective oxide thickness of 50 at% PtPd samples determined using APT at various quenched temperatures during thermal cycling, (ii) the NO conversion rate measured by FBR with FT-IR. Both sets of data were obtained under a same gas atmosphere of 500 ppm NO and 3% $O_2$ in $N_2$.*

During the cooling period from 200 °C to 80 °C and at the onset of the following heating cycle, a significant reduction of the oxide layer occurs, evidenced by a decrease in thickness to 0.59 nm, as shown in Figure 9(i). Since this reduction does not completely eliminate the oxide, only a partial recovery of catalytic activity is achieved, which corresponds to the observed moderate increase in NO conversion. This behavior clearly illustrates the reversible and dynamic nature of surface oxidation and reduction processes in the PtPd catalyst under operational conditions, emphasizing how temperature variations govern transient changes in catalytic performance.

## 3.5 Reversible oxidation of catalyst surface

The results obtained from both Pt and PtPd catalysts reveal remarkably similar dynamic changes in the surface oxide layers during NO oxidation. This similarity is closely linked to the observation of inverse hysteresis in NO conversion across both catalytic systems. Particularly noteworthy is the partial restoration of NO conversion during the first cooling and subsequent second heating ramps, especially at lower temperatures (T< 200°C). This phenomenon is directly correlated to the measurements of oxide layer thickness, which demonstrate a partial reduction of the oxidized catalyst surface under these conditions. The current observation significantly diverges from the results obtained for pure Pd in our earlier work [29], as presented in Figure 10. The pure Pd catalyst exhibited only a one-time conversion during the initial heating process. This was followed by a constantly reduced conversion rate throughout all subsequent temperature ramps, with no recovery of catalytic activity. This lack of recovery is attributed to the pure Pd being fully oxidized during the first heating phase, and the resulting oxide layer remaining stable throughout the following thermal cycles.

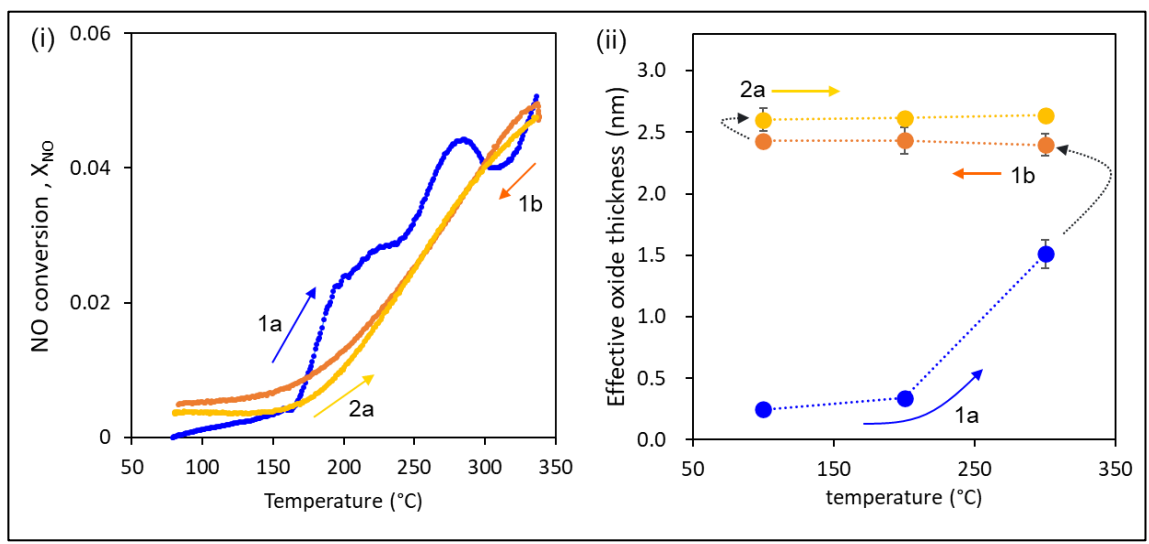

*Figure 10. Diagram of (i) the NO conversion rate measured by FBR, (ii) the effective oxide thickness of pure Pd samples determined using APT, all under a gas atmosphere of 500 ppm NO and 3% O₂ in N₂ [29].*

Building on these observations, the possible mechanisms possibly explaining the dynamic behavior of Pt and PtPd in contrast to pure Pd must be discussed, specifically addressing why such partial recovery of NO conversion emerges in the presence of NO at lower temperatures. For the Pt system, three key reactions are considered. Their Gibbs energy change ($\Delta G = \Delta H° - T\,\Delta S°$) was calculated as shown in Table 3. The thermodynamic parameters, standard enthalpy ($\Delta H°$) and entropy ($\Delta S°$) of formation, were derived from datasets available in the NIST Chemistry WebBook [48] and the work of C. Wang et al.[49].

$$NO + \frac{1}{2} O_2 \rightleftharpoons NO_2 \qquad \text{Reaction 1}$$

$$Pt + \frac{1}{2} O_2 \rightleftharpoons PtO \qquad \text{Reaction 2}$$

$$PtO + NO \rightleftharpoons Pt + NO_2 \qquad \text{Reaction 3}$$

*Table 4. Gibbs free energy change of formation for NO oxidation, Pt oxidation and PtO reduction*

| T (°C) | T (K) | Reaction 1 ΔG (kJ/mol) | Reaction 2 ΔG (kJ/mol) | Reaction 3 ΔG (kJ/mol) |
|---|---|---|---|---|
| 25  | 298 | -35.35 | -34.49 | -19.72 |
| 100 | 373 | -29.85 | -20.53 | -28.19 |
| 200 | 473 | -22.52 | -1.90  | -39.49 |
| 300 | 573 | -15.19 | 16.73  | -50.79 |
| 400 | 673 | -7.86  | 35.35  | -62.08 |

Experimental observations reveal that within the temperature interval of 200–300 °C, Pt oxidation readily occurs in the presence of oxygen. At higher temperatures (>350 °C), however, the reduction of Pt oxide becomes thermodynamically favored [2] [3] [8] [20]. This observation aligns with the Gibbs free energy values listed in Table 3, confirming that PtO instability increases at elevated temperatures.

In contrast, when the temperature drops below approximately 200 °C, the introduction of NO promotes the reduction of PtO (Reaction 3), restoring the metallic Pt state. This phenomenon, crucial for the observed reactivation of catalytic performance, emerges from the interplay between thermodynamic driving forces and kinetic constraints. Although the NO-to-$NO_2$ equilibrium favors the product side even at low temperatures—as shown in standard equilibrium plots [3] — the actual reaction rate remains limited by the low catalytic activity of the platinum oxide formed during the preceding cooling stage (350–250 °C). Yet, under these oxygen-lean and NO-rich conditions, the strong thermodynamic potential combined with the abundant availability of NO effectively drives the reduction of PtO to metallic Pt, with NO simultaneously oxidized to $NO_2$. This NO-assisted reduction accounts for the partial recovery of catalytic activity observed at lower temperatures.

As explained by Hauff et al.[3], both the formation and reduction of platinum oxide are intrinsically slow processes that evolve over minutes to hours. They identified the oxidation by $O_2$ and reduction by NO as competing reactions, showing that the activation energy of Pt oxidation (Reaction 2) is higher than that of PtO reduction (Reaction 3). While platinum oxidation is thermodynamically feasible, its low-temperature rate is hindered by an elevated activation barrier; once sufficient thermal energy is provided, oxidation surpasses reduction, underscoring the dominant influence of kinetics over thermodynamics.

Further insight into this complex redox mechanism is obtained in the APT analysis, which elucidates nanoscale oxidation behavior complementary to kinetic observations. In a related study on pure Pd tips (Y. Lee et al.[29]), the previously described oxygen cluster ions (e.g., $^{64}O_4^+$) appeared in the mass spectra only during the initial heating phase (1a) of NO oxidation. Once a stable PdO layer had formed, these cluster peaks disappeared, replaced by $PdO_x$ molecular species. This strongly indicates that oxygen clusters in the APT spectra are a distinctive characteristic of the earliest stages of physisorption, chemisorption, or oxide nucleation, whereas metal-oxygen molecular ions safely indicate the presence of a well-established oxide phase.

In the case of Pt, these oxygen cluster peaks (e.g., $^{48}O_3^+$, $^{64}O_4^+$, $^{80}O_5^+$) were even observed not only during the cooling period (1b) but also reappeared during the subsequent heating ramp (2a). The persistent detection of oxygen clusters, instead of stable $PtO_x$ species, throughout these temperature cycles suggests that while surface oxidation occurs, a stable, bulk-like PtO layer neither fully develops nor remains under such dynamic conditions. Instead, the Pt surface exists in a transiently oxidized state, fluctuating among chemisorbed oxygen, incipient oxide nuclei, and metallic regions.

One of the key characteristics of the PtPd alloy is its inherent tendency for Pd atoms to segregate toward the surface. This observation leads to a natural question: if the surface enriches in Pd so that Pd becomes the majority component is overwhelming, could a stable oxide film develop similar to that on pure Pd? Interestingly, the PtPd APT analysis demonstrates a reversible redox behavior that resembles the response of pure Pt, rather than exhibiting the persistent oxide typically observed at Pd surfaces. As shown by the effective oxide thickness measured by APT, the oxide layer on the PtPd alloy after Pd segregation exhibits an intermediate thickness—greater than that on pure Pt but still thinner than that on pure Pd. More importantly, the oxidized catalyst surface undergoes reduction at lower temperatures, even in the presence of surface-enrichment in Pd. This behavior is further supported by the consistent detection of oxygen cluster peaks ($^{48}O_3^+$, $^{64}O_4^+$, $^{80}O_5^+$, and even $^{128}O_8^+$) in the APT mass spectra, suggesting that oxidation remains confined to the nucleation stage and does not progress into a fully coherent oxide.

This unexpected oxidation behavior can be attributed to the intrinsic interactions between Pt and Pd within the alloy. Pt, known for its sluggish oxidation kinetics, likely inhibits the formation and stabilization of a continuous PdO phase. In addition, the mixed Pt–Pd surface may exhibit modified electronic properties that reduce affinity to oxygen and suppress the growth of thick oxide layers. As a result, the surface remains in a dynamic redox state, which facilitates enhanced NO oxidation at lower temperatures and very likely accounts for the inverse hysteresis observed in the PtPd catalyst.

## 4. Conclusion

In this study, the surface evolution of Pt and PtPd catalysts - both widely used in diesel oxidation catalysts (DOCs) - was investigated under NO oxidation conditions, specifically during a temperature ramping in a gas mixture containing 0.05% NO and 3% $O_2$ in $N_2$. The study focused on how surface reactions relate to the observed NO conversion behavior. Notably, both Pt and PtPd catalysts exhibited inverse hysteresis in NO conversion, and the APT-based measurements of effective oxide thickness revealed similar surface transformation patterns for both catalyst systems. The findings confirm that both metal catalyst surfaces undergo a reversible oxidation redox process. A close correlation with the gas conversion rate strongly suggests that the reversible surface oxidation is responsible for the observed inversed hysteresis in gas conversion.

For the pure Pt catalyst, surface oxidation during heating is clearly demonstrated to a maximum effective oxide thickness of 0.57 nm. However, when cooling below a temperature of 200 °C the oxide thickness decreased from 0.57 nm to 0.35 nm. This behavior is due to three factors: Firstly, the sluggish oxidation kinetics hinders the development of a stable bulk Pt oxide phase, leaving the platinum surface predominantly in state of chemisorbed oxygen. Second, Pt oxide decomposes at high temperatures due thermodynamic destabilization. Third, at low temperatures PtO is reduced in NO containing atmosphere. The observations suggest that the redox behavior of Pt is majorly governed by reaction kinetics rather than by thermodynamic stability. In particular, the activation energy for Pt oxidation in oxygen is likely higher than that for PtO reduction in NO. As a result, Pt oxidation is suppressed at low temperatures, while enhanced oxidation occurs at intermediate temperatures due to increased thermal activation.

For the PtPd alloy, a similar redox behavior is observed. The most significant additional feature is the compositional change of Pt and Pd on the surface. Under oxidizing conditions, Pd atoms segregate to the surface. This phenomenon begins above 300°C during the first heating cycle. For instance, the initial composition of 50 at% Pt and 50 at% Pd shifted to 31 at% Pt and 69 at% Pd at the 1b-200 °C stage. The segregation phenomenon is supported by the thermodynamic preference for Pd oxidation (lower Gibbs free energy) and electronic structure effects (e.g., a shift in the d-band center promoting stronger oxygen binding).

In general, the PdPt alloy demonstrates thicker oxide layers than pure Pt. However, despite the segregation induced enrichment of Pd, the PtPd alloy still does not form a stable oxide layer in contrast to studies of pure Pd. Instead, the alloy exhibits reversible redox behavior, closely resembling that of pure Pt. The dynamic surface transformation during the temperature ramping showed the same partial reduction in the first cooling and second heating phases. This behavior aligns well with the inverse hysteresis observed in the NO conversion of the PtPd catalyst.

The fact that oxide films formed at Pt and PtPd alloy surfaces are less stable than those formed on pure Pd is further corroborated by (i) significant local variation in the surface oxygen content and (ii) the appearance of unusual oxygen cluster molecules in the mass spectra, which are demonstrated to be a reliable indicator for a transient oxidation states.

In conclusion, the significance of this study lies in its ability to uncover detailed surface transformations in nanoparticle-based catalysts—phenomena that are challenging to characterize experimentally. By employing atom probe tomography (APT) on nanoscale tips, our research provides high-resolution insights and establishes a robust analytical methodology for probing surface and compositional changes at the atomic scale.

The study not only deepens the fundamental understanding of oxidation behavior in noble metal alloys, but also offers strategic insights for the rational design of next-generation alloy catalysts with enhanced durability and performance under dynamic and fluctuating reaction environments. Understanding how Pt and Pd atoms dynamically rearrange under operating conditions—such as Pd surface segregation and oxide layer formation—enables predictive control over the active surface state. Importantly, the reversible nature of oxide formation and the partial recovery of catalytic activity highlight pathways for designing catalysts that resist irreversible deactivation. For instance, suppressing the formation of thick and stable oxide layers directly contributes to improved catalyst lifetimes.

**Acknowledgement**

We are grateful to the financial support by the Deutsche Forschungsgemeinschaft (DFG) under the collaborative project SCHM 1182/18-1 / TU 440/1-1 and within the collaborative research center CRC1333/C03 (grant 358283783). Also, the investments of the utilized atom probe and the SEM/FIB instrument were jointly supported by the Deutsche Forschungsgemeinschaft and the Baden-Württemberg-foundation, which is gratefully acknowledged.